# Radial Transport in the Solar Nebula: Implications for Moderately Volatile Element Depletions in Chondritic Meteorites

Fred J. Ciesla
Department of Terrestrial Magnetism
Carnegie Institution of Washington
5241 Broad Branch Road, NW
Washington, DC 20015
ciesla@dtm.ciw.edu
voice: (202) 478-8482
fax: (202) 478-8821

## Abstract

In this paper, the possibility that the moderately volatile element depletions observed in chondritic meteorites are the results of planetesimals accreting in a solar nebula that cooled from an initially hot state (temperatures > 1350 K out to ~2-4 AU) is explored. A model is developed to track the chemical inventory of planetesimals that accrete in a viscously evolving protoplanetary disk, accounting for the redistribution of solids and vapor by advection, diffusion, and gas drag. It is found that depletion trends similar to those observed in the chondritic meteorites can be reproduced for a small range of model parameters. However, the necessary range of parameters is inconsistent with observations of disks around young stars and other constraints on meteorite parent body formation. Thus, counter to previous work, it is concluded that the global scale evolution of the solar nebula is not the cause for the observed depletion trends.

Pages: 31
Figures: 9
Tables: 1
Words: 11,925

# Introduction

Chondritic meteorites have long been recognized to be relatively unaltered products of our protoplanetary disk, the solar nebula. The relative abundances of the elements they contain closely match that which is observed in the photosphere of the sun (e.g. Lodders 2003). The meteorites as a whole, as well as their individual components —chondrules, refractory inclusions, metal grains, and matrix—record a wide variety of chemical and physical environments that existed within the solar nebula. Understanding what these environments were and how they formed in the solar nebula is a challenge, and must be done in the context of our understanding of how protoplanetary disks evolve. Doing so will provide insight as to what the interiors of other protoplanetary disks are like, which would be valuable as these regions, particularly at small heliocentric distances, cannot be directly probed by telescopic observations.

One of the longstanding, and as of yet, unsolved mysteries in the study of these meteorites is the cause of the depletion of moderately volatile elements (MOVEs, those that condense between ~650-1350 K under solar nebula conditions). This has been the focus of many research efforts over the last four decades (Anders 1964; Larimer and Anders 1967; Wasson and Chou 1974; Wai and Wasson 1977; Wasson 1977; Palme 1988; Palme and Boynton 1993; Cassen 1996; 2001; Alexander 2005; Bland et al. 2005; Yin 2005). Specifically, the depletion of these elements is correlated with condensation temperature--that is, the relative abundance of an element with a lower condensation temperature is less than that of an element with a higher condensation temperature. The depletion trends for the CM, CO, and CV meteorites are shown in Figure 1. The trends are to first order monotonic, with each chondritic type exhibiting a unique depletion trend (except CI chondrites for which are depleted in only the most volatile elements when compared to the composition of the solar photosphere). Deviations from the monotonic trend can be attributed to uncertainties in the relative abundances or condensation temperatures of the elements, or they may be due to the actual evolution of the materials in the solar nebula (Cassen 1996). Here the focus is on understanding the first order depletion trend, which is shared by the different meteorite types.

Anders (1964) originally attributed the MOVE depletion trend as due to the mixing of two different meteoritic components: a volatile-rich component (matrix) and a volatile-poor component (chondrules). Later, Wasson and coworkers (Wasson and Chou 1974; Wai and Wasson 1977; Wasson 1977) suggested instead that the trend was due to incomplete condensation of the nebular gas that was dissipated as it cooled, meaning that the inventory of volatile elements in the gas would diminish over time and there would thus be less of a given element once the gas had cooled below the corresponding condensation temperature. More recently, it has been suggested that the MOVE depletion was inherited from the natal molecular cloud from which the solar system was born, as the gas in molecular clouds are depleted in refractory elements that have condensed out as grains (Yin 2005).

While the exact cause for the observed trend has been hotly debated and continues to be so today, the incomplete condensation model of Wasson and coworkers has generally been the favored explanation (Bland et al. 2005). Specifically, Palme et al. (1988) and Palme and Boynton (1993) identified a number of reasons why the two-component model failed to explain the observed trend, including the fact that chondrules contain a significant amount of volatiles and therefore are not "volatile depleted" as required by the Anders model. Palme

(2001) also argued that the Rb/Sr inventory of chondritic meteorites is inconsistent with the molecular cloud inheritance theory, though Yin (2005) outlines counterarguments.

In addition to the chemical arguments, the incomplete condensation theory gained a significant amount of support by the fact that astrophysical models of solar nebula evolution were able to qualitatively reproduce the observed depletion trends (Cassen 1996; 2001). In Cassen's models, the elemental inventory of planetesimals were tracked as they formed in a cooling nebula that evolved as mass was transported inward and accreted by the young sun and angular momentum was transported outward to compensate. As the nebula cooled at a given region, the elements in the gas phase with condensation temperatures above the local temperature were incorporated into the planetesimals that were present. However, because the cooling was associated with mass loss from the disk, the gas was constantly removed, meaning that the more volatile species were depleted in comparison to the elements with higher condensation temperatures. In order for this trend to be observed in elements such as Si, with a condensation temperature of ~1350 K, the chondrite formation region (out to ~4 AU) had to have initially been above this temperature. This would imply the nebula was initially very massive and compact to achieve such high temperatures at these heliocentric distances. Such a situation is thought to be consistent with the uniform isotopic composition (except for those of oxygen) of chondritic materials (Palme 2001).

In this paper, a new model is developed to evaluate whether the incomplete condensation theory for the origin of the MOVE depletion could be due to the dynamical evolution of the solar nebula. This new model is similar to those used to describe the structure and evolution of protoplanetary disks observed around young stars, and tracks the radial migration of solids and vapor due to advection, diffusion, and gas drag. In the next section the details of the model used here are described and are directly compared to those used by Cassen (1996; 2001). Results of the new model and the implications for the compositions of meteoritic components and planetesimals are discussed in the sections that follow. Issues that have not been considered in previous models for the MOVE depletion are then presented, with a summary and ideas for future work outlined at the end.

## Solar Nebula Model

The nebula models developed by Cassen (1994; 1996; 2001) were motivated by the desire to understand chondritic meteorites and materials in the context of astrophysical processes similar to those observed or inferred to occur in other disks around young stars. Specifically, these models were used to explain the chemical and isotopic properties of solar system materials as the result of their dynamical evolution in a protoplanetary disk. The disk began as a hot, compact object which cooled over time as the disk thinned as a result of mass and angular momentum transport. The rate of mass transport also diminished with time, leading to less thermal energy being generated at later stages of evolution. In addition, solids coagulated into larger objects, reducing the opacity of the disk, allowing radiative energy to escape more easily, which also aided in the cooling of the disk. As these solids grew in the disk, they decoupled from the gas, and their dynamical evolution differed from that of the gas. Cassen's models tracked the different environments that solids would be exposed to during the lifetime of the nebula and compared these results to what was observed in meteorites. As such, these models represent major steps towards linking the fields of meteoritics and astrophysics.

The work presented here builds off of Cassen's pioneering work to further investigate how the MOVE depletion trend may have developed in the solar nebula. The evolution of the solar nebula is described by a similar set of equations as those used to describe disks around young stars. While many of the treatments here are similar to, or based on, those of Cassen (1996; 2001), new effects are considered. The reasons for considering these effects and how they are treated in the model are described below. The model used here is based on the model of Ciesla and Cuzzi (2006), which tracked the dynamical evolution of water-bearing species in a viscous protoplanetary disk. A qualitative discussion of the model and the key differences between it and those of Cassen are described below. The interested reader is directed to Ciesla and Cuzzi (2006) for the details of the model and specific description of how the calculations are performed.

**Disk Evolution**

To describe the mass and angular momentum transport, the disk is assumed to have a viscosity given by $\nu=\alpha c H$ at each location in the disk, where $c$ is the local speed of sound, $H$ is the disk scale height, and $\alpha$ is the turbulence coefficient (a free parameter) which represents the level of turbulence in the disk and is assumed to have a value less than 1 (Shakura and Sunyaev 1973; Lynden-Bell and Pringle 1974; Ruden and Pollack 1991; Stepinski 1998). Because of the viscosity, shear stresses develop between neighboring annuli in the disk that are responsible for driving the evolution of the disk. The orbital velocity of the gas falls off as $\sim r^{-1/2}$, thus at a given location in the disk, the gas immediately inward will have a slightly larger velocity than the gas immediately outside that location. As a result, the outer annulus exerts a stress that works to slow down the gas in the inner annulus, causing it to lose angular momentum and therefore move inward, towards the central star. These same stresses work to accelerate the gas in the outer annulus, causing it to gain angular momentum, and therefore move outward, away from the central star. It is these stresses that cause the disk to lose mass to the central star and grow in radial extent over time.

This treatment differs from the approach taken by Cassen (1996; 2001), who defined an evolutionary timescale, $t_e$, which determined the rates of mass and angular momentum transport. This was done to avoid having to define a process that would be responsible for driving mass transport in protoplanetary disk. This driving force remains the subject of ongoing research today. As such, the use of the $\alpha$-viscosity does not necessarily represent an improvement, as the uncertainty in what determines the value of $\alpha$ is still debated. The most promising candidate is the magnetorotational instability or MRI (Balbus and Hawley 1991). In order for the MRI to operate, however, the gas must be sufficiently ionized to couple to local magnetic fields, which happens only at the extreme inner regions of the disk where temperatures are in excess of ~1000 K such that ionization is achieved through molecular collisions, and in the very outer parts of the disk and in the upper surface layers where X-rays and cosmic rays penetrate and are absorbed resulting in photoionization (Gammie 1996). The midplane region of the solar nebula in the planet formation region (<~10 AU) may not have been ionized to sufficient levels to allow for the MRI to operate, meaning it would be laminar and no mass or angular momentum would be transported through this "dead zone." However, Fleming and Stone (2003) found that it is possible for turbulence to be "stirred up" in the dead zone by the MRI-active layers at high altitudes. Thus, while the MRI may not be present, turbulence may still exist at the midplane of the disk, allowing mass and angular momentum transport to take place. To address the uncertainties surrounding this issue, a range of values is investigated for $\alpha$, which is assumed to be constant throughout the disk for simplicity.

The surface density evolution of a viscous protoplanetary disk is described by the equation:

$$\frac{\partial \Sigma}{\partial t} = \frac{3}{r}[r^{1/2}\frac{\partial}{\partial r}(r^{1/2}\nu\Sigma)]$$

where $\Sigma$ is the surface density of the disk (mass per area), $r$ is the heliocentric distance, and $\nu$ is the viscosity as described above. The viscosity is dependent on the local speed of sound, and therefore, the local, midplane temperature, $T_m$, which is given by:

$$T_m^4 = \frac{3}{4}\tau T_e^4$$

where $\tau$ is the optical depth from the midplane of the disk to the surface, and $T_e$ is the temperature of the disk surface, which is set by balancing the energy gained through viscous dissipation, stellar illumination, and absorption of radiation form an ambient field with the energy lost through radiation. The optical depth is given by $\tau=\kappa\Sigma_d/2$, where $\kappa$ is the opacity of the solids, taken here to be 1000 cm$^2$/g, which in a gas of solar composition gives an opacity of the gas equal to 5 cm$^2$/g, roughly equal to the value found by Pollack et al. (1994) for a solar composition gas at pressures and temperatures expected in the terrestrial planet region of the solar nebula. A similar value was used by Cassen (1994; 1996; 2001). The factor of $\Sigma_d$ represents the surface density of dust particles, meaning that as the dust is removed through coagulation, the optical depth decreases, and the interior of the disk will cool.

**Growth of Solids**

The solids considered here are assumed to be predominately Mg, Si, and Fe minerals, particularly forsterite and metal grains. These minerals constitute the majority of mass in chondritic meteorites (generally > 90%), and the condensation of MOVEs takes place when those elements react with these solids. While water ice can make up a significant fraction of the solids in the cool regions of the disk, its condensation temperature (~160 K) is well below the temperature range of interest for MOVE condensation (~650-1350 K). Thus the dynamics of water ice are not considered in this model.

Throughout the time that the disk evolves, the solids contained within it grow through collisions. Cassen (1996; 2001) considered two different sizes of solids: *dust* and *planetesimals*. Solids were initially uniformly dispersed as dust, and Cassen (1996; 2001) defined a coagulation timescale which set the rate at which planetesimals grew from the available dust at a given location of the disk. Here, as in Ciesla and Cuzzi (2006), a third species of solids is considered, *migrators*, which represents the size range of solids that fall between the two extremes. This third set of solids is so named as they are strongly affected by gas drag migration, which can be a major mechanism for redistributing elements in a protoplanetary disk.

To account for the particle growth, a similar approach as used in Cassen (1996; 2001) is adopted in that timescales are still set which determine the rate at which larger bodies form from smaller ones. Here, however, two timescales are needed. The coagulation timescale, $t_c$, defines the rate at which migrators form from the dust population, and the accretionary timescale, $t_a$, defines the rate at which planetesimals form from the migrator population. The timescales are assumed to vary with location as in the Cassen models in that:

$$t_a(r)=t_a(1\ AU)\Omega(1\ AU)/\Omega(r)$$

where $\Omega$ is the local Keplerian orbital period at a given location in the disk. The same relation is used for the coagulation timescale.

Values for $t_c$ and $t_a$ can be estimated from numerical studies of coagulation and planetesimal formation. Cassen (1996; 2001) estimated that planetesimals would form from a dust population on timescales on the order of $10^3$-$10^4$ years. Beckwith et al. (2000) suggested that the formation of the migrator population from the dust population would occur on timescales of the order $10^4$ years and that planetesimals would form from the migrator population on timescales of another $10^4$ years. These represent only rough estimates as disks around young stars remain optically thick for millions of years (Haisch et al. 2001). As a result, a wide range of values is considered in this work for the different growth timescales.

Figure 2 compares the two-stage growth of planetesimals used here to the single-stage used by Cassen (1996; 2001). Plotted are the surface density distributions of all the solids at a given location in the nebula, assuming a quiescent (non-evolving) disk with no radial transport. In all cases, the timescales are assumed to have values of $10^4$ years. The rate of change of the dust population is the same in each case as the coagulation timescales (dust depletion timescales) are the same. However, in the two-stage model of planetesimals growth, the dust is incorporated into migrators, which are then incorporated into planetesimals over time. This slows the formation of planetesimals in comparison, as can be seen in the figure.

**Radial Transport of Material**

As the gas, predominately $H_2$ and He, in the disk is transported due to the viscous stresses described above, the material suspended in the disk will be moved about as well. Dust particles are so small that they tend to be coupled to the gas and therefore behave dynamically the same as the gas. More specifically, this is true for solids with relatively short stopping times, $t_s$, which is given by:

$$t_s=a\rho/\rho_g c$$

where $a$ is the radius of the particle, $\rho$ is the mass density of the particle, and $\rho_g$ is the local gas density in the disk. This relation holds for those particles whose radius is less than the mean-free path in the gas ($a<\sim 10$ cm for the inner disk). When the stopping time is less than the orbital period ($t_s\Omega<1$) the particle will generally follow the motions of the gas. Cassen (1996; 2001) accounted for this by determining the net advective flow of the gas at each location in the disk based on the rate of mass transport, either inward or outward depending on the location, and allowed the dust (or vapor if above the condensation temperature) to move with that same velocity. Here, the dust (and vapor) dynamics are described by the same equation that describes the dynamics of the disk gas, meaning that they also follow the large-scale flows associated with disk evolution, moving inward or outward in accord with how mass and angular momentum are transported.

Planetesimals are assumed to be large enough that $t_s\Omega \gg 1$. This means that the stopping times are so large that the planetesimals are relatively unaffected by the gas. Thus once planetesimals form, they are assumed to remain at the location of their formation. This was also assumed to be the case by Cassen (1996; 2001).

The major difference in the handling of the solid bodies between this model and that of Cassen (1996; 2001) is in the way that the radial transport of the migrators, which are those bodies with $t_s\Omega \sim 1$, is accounted for. These bodies are strongly affected by gas drag migration which arises due to the fact that the gas in the disk is partially supported against gravity by a radial pressure gradient, causing the gas to orbit the central star at a less than Keplerian rate (Weidenschilling 1977). As a result, solids in the disk move through the gas with a relative velocity, and thus feel a drag force that impedes their orbital motion. The bodies thus lose energy and momentum to the gas, and migrate inwards over time. Those bodies with $t_s\Omega = 1$ are most strongly affected, and migrate inwards at rates of ~1 AU/century, meaning a meter-sized body would migrate from 10 AU to the inner edge of the disk in 1000 years, which is a small time compared to many other timescales of interest in protoplanetary disk evolution.

To account for the redistribution of material by gas drag, Cassen (1996; 2001) assumed some fraction of solids was lost at every time interval to migration into the sun, and removed the corresponding amount of solids from the nebula over time. Recent work, however, has demonstrated that very few migrators actually survive their transit to the sun and thus tracking their detailed dynamical evolution is critical to understanding the chemical evolution of the solar nebula. Cuzzi et al. (2003) showed that the influx of silicate and carbon-rich migrators to inside the silicate evaporation front (where temperatures exceed the condensation temperature of forsterite, taken to be ~1350 K) could create an environment that allowed Calcium-Aluminum-rich Inclusions (CAIs) to form over a period of a few hundred thousand years. This idea was expanded upon by Cuzzi and Zahnle (2004) who envisioned evaporation fronts for all major chemical species and predicted that the inward flux of migrators would be high during the early stages of evolution and would lead to species becoming enhanced in the vapor phase inside their evaporation front and then depleted as the inward flux diminished over time. Ciesla and Cuzzi (2006) explicitly modeled this evolution for water, and found that the timing and magnitude of the concentration fluctuations inside the snow line (the water ice evaporation front) were sensitive to the assumed disk structure and that the mass of the disk beyond the snow line would severely limit the level to which the inner disk was enhanced in water vapor during the early stages of evolution. The inward migration of water ice has also been invoked as a way of carrying $^{17}O$ and $^{18}O$ released from CO molecules as a result of self-shielding processes to the region where chondritic materials formed, explaining the observed oxygen isotope evolution in these meteorites (Yurimoto and Kuramoto 2004; Lyons and Young 2005; Krot et al. 2005). Ciesla and Lauretta (2005) also demonstrated that the sluggish dehydration kinetics of phyllosilicates would allow these minerals to be transported from the outer asteroid belt to the region where the Earth accreted by gas drag migration, even though they would migrate through environments that exceeded their formation temperature. This would allow the planetesimals there to acquire a minor amount of hydrated minerals, offering a possible explanation for the presence of water on Earth. Thus the dynamics of the migrators must be taken into account when considering the chemical evolution of the solar nebula.

In addition to gas drag, the migrators are redistributed by the large-scale flows associated with disk evolution. Their evolution due to these flows is described by the same equation as that which describes the evolution of the nebular gas and dust, but the viscosity is given by:

$$\nu_m = \nu/(1+t_s\Omega)$$

or $\nu_m = \nu/2$ for the specific case considered here, meaning that these larger objects are less coupled to the gas than the smaller bodies due to their larger stopping times (Cuzzi and Weidenschilling 2006). (The value of $t_s\Omega$ is also referred to as the Stokes number, $St$, in a turbulent disk, as the largest eddies at a given location are expected to overturn on timescales comparable to $\Omega^{-1}$.) As a result of gas drag, these bodies also move inwards at a rate given by (Weidenschilling 1977):

$$v_r = \frac{\pi}{\rho_g}\left(\frac{r}{GM}\right)^{1/2}\frac{dP}{dr}$$

where $M$ is the mass of the sun and $dP/dr$ is the local pressure gradient in the disk.

In reality, the inward migration rate of bodies is a strong function of size, with the maximum rate given by bodies with $t_s\Omega=1$. Thus, here it is assumed that bodies move inwards at the maximum rate. The actual *mass flux* of material is what is important, and this will be set by the combination of inward migration rate and the surface density of migrators. The surface density of the migrator population is determined by the interplay of dust coagulation and planetesimal accretion, and therefore the timescales that these processes operate on are what control the inward mass flux. Thus, while the migrating population is intended to include bodies which migrate inwards at slower rates than identified above, the different mass fluxes that would arise as a result are accounted for by considering a range of coagulation and accretionary timescales.

**Evolution of Elemental Concentrations**

While the various solids are redistributed due to the processes described above, they can carry trace species such as the moderately volatile elements with them. The kinetics of chemical reactions, while possibly important, are ignored in this work. The effects that non-instantaneous reactions may have are discussed in detail below. Following Cassen (1996; 2001), it is assumed that an element is located in the solids when the local temperature is below the condensation temperature of that element and released to the vapor when solids reach an area that exceeds the condensation temperature.

Allowing immediate release of the element to the gaseous phase is reasonable if the element has an unimpeded path from the solid containing it. This may not be the case when an element is located at the center of a migrator and must make its way through ~1 meter of material before being released to the nebula gas. If the body was porous, then the element would likely escape more easily than if the body was more consolidated. Two extreme cases are considered here: the case where the release of an element from a migrator is prohibited by the solids in the body and therefore released only when the silicates are vaporized, and the case where the release of the element is not inhibited and therefore is incorporated into the gas as the migrator crosses the corresponding evaporation front.

Specific elements are not considered here, rather hypothetical elements with condensation temperatures of 1200, 1000, 800 and 600 K are tracked. As the MOVE depletion trends that are the focus of this work simply correlate with temperature, this treatment is well justified. These elements add negligible mass to the solids, which are predominately composed of Mg-rich silicates and Fe metal. When the element vapor diffuses outwards in the disk to regions below its condensation temperature, it is assumed to condense on the silicate dust present there, regardless of how much dust is available—the specific reaction pathways are not considered. This treatment likely is not a significant stretch as silicate dust can diffuse outwards along with the element vapor and in no cases is the silicate dust significantly depleted with respect to that vapor, meaning that the condensation should not be limited by available reaction sites.

## Results: Planetesimal Compositions

Each model that was investigated was defined by a set of parameters that determined the initial disk structure, the turbulence parameter, as well as the rates at which solids grew within the disk. The initial disk structure is assumed to be described by the relation:

$$\Sigma(r<R_0) = \Sigma(1\ AU) \left(\frac{r}{1\ AU}\right)^p$$

where $R_0$ represents the initial radius of the disk. Once the initial structure of the disk was established, the disk was allowed to evolve viscously as determined by the value of $\alpha$ used. While the disk evolved, the migrators and planetesimals formed as described above using the timescales given by $t_c$ and $t_a$. A wide parameter space was explored to understand how different disk structures and evolutionary rates could lead to changes in the moderately volatile element depletion in those planetesimals that formed at various locations in the disk. Here, only a small subset of cases is presented, to illustrate the possible outcomes that could be produced and how the outcomes vary with different parameter choices. The parameters for the different cases presented here are given in Table 1.

The first case described is analogous to those simulations done by Cassen (1996; 2001) where dust was incorporated rapidly into planetesimals (the lifetime of migrators, and thus the transport associated with their presence is suppressed by the short accretionary timescale). The evolution of the disk and the depletion trends that result can be seen in Figure 3. These parameters were chosen to illustrate that, despite using a slightly different model for the nebular evolution, the model here produces results similar to those found by Cassen (1996; 2001). Specifically, in the inner disk, there is a general trend of the relative abundances of elements contained in planetesimals decreasing with lower condensation temperatures. The initial temperature of the disk at a particular location defines the temperature at which the depletion begins. That is, when planetesimal formation begins at a given location, those planetesimals incorporate all those elements that have locally condensed. For example, if a planetesimal forms in a region initially at T=800 K, it will begin with a full inventory (CI-like abundance) of elements with condensation temperatures greater than 800 K. As the disk evolves due to mass loss and the decrease in opacity associated with dust coagulation, the local temperature drops (Panel A), allowing solids to become more volatile enriched with time. However, the amounts of these more volatile species that are available to be accreted diminish with time as a result of this evolution. Thus, there is a relative depletion of these more volatile species (Panel B). In order to reproduce the MOVE depletion trend, which begins with Si, the temperature at a given location of the disk must have exceeded the forsterite condensation temperature ~1350 K. Such high temperatures are only reached inside of ~2 AU in this particular model, meaning that the complete MOVE depletion trend is reserved only for those bodies that formed interior to this distance.

In Cases 2 and 3, the same disk structure is used as in Case 1, but now transport via gas drag is allowed, setting the accretionary timescale at 1 AU to be $10^3$ years. In Case 2, it is assumed that vaporizing species are able to easily escape from the migrators, whereas in Case 3 it is assumed that the species are not released to the gas on timescales less than the dynamical timescales. There is little change in the evolution of the disk surface density or temperature structure in these cases versus that in Case 1. The reason for this is that the only way in which solids affect the evolution of the disk in a viscous disk is by determining the optical depth, which plays a role in determining the temperature of the disk and the local viscosity (in an α-viscosity model). As such, the only way that the distribution of dust in the disks vary from that in Case 1 is by the inward transport of the migrators across the silicate evaporation front, where they vaporize and the resultant vapor diffuses outwards to condense as dust. This only has a significant effect near the silicate evaporation front, and because of the relatively short accretionary timescale does not lead to drastic changes in the evolution of the disk. Longer accretionary timescales would allow for greater transport, and thus, greater shifts in the dust distribution.

The bulk compositions of the planetesimals that form in Cases 2 and 3 are strongly affected by the inward transport of elements in migrators. In the case of rapid escape of a given element from the migrators (Case 2), the concentration of that element is enhanced in the vapor phase inside its evaporation front, similar to what is expected for water inside the snow line (Cuzzi and Zahnle 2004; Ciesla and Cuzzi 2006). As the disk cools, the element condenses locally at a slightly higher concentration than solar, meaning more of it is incorporated into the planetesimals than in the case where gas drag migration is inhibited. A depletion trend is still evident at small heliocentric distances, though it is not as steep as in Case 1.

Essentially no depletion trend develops if it takes long periods of time for elements to diffuse outwards from the migrators as found in Case 3 (Figure 5). This is because the migrator population at a given location is dominated by migrators that formed at greater heliocentric distances. This is due to the rapid rate at which these bodies move inwards over time—for the accretionary timescales of interest here, a migrator can travel ~10 AU before being incorporated into a planetesimals. Thus, the composition of the planetesimals that accrete in this scenario are not representative of the material that forms at that location, but rather, more representative of those materials that form further out in the disk.

Based on the model results here, it is possible to produce depletion trends that are similar to the MOVE depletions observed in chondritic meteorites and similar to the results found by Cassen (1996; 2001), even when allowing for the redistribution of materials by gas drag. However, it is not a robust result. Certain conditions must be met in order for the depletion trends to be produced in bodies located in what would become the current day asteroid belt (2-4 AU). Firstly, because all moderately volatile elements must start in the vapor phase, this requires temperatures above ~1350 K beyond at least 2 AU. Secondly, in order for the planetesimals to develop a chemical "memory" of the cooling history of the disk, they must accrete rapidly so that they can preserve significant amounts of material from every temperature interval on the cooling curve. In all of the cases shown here, the depletion trends are plotted after 500,000 years. Beyond this time the temperatures in the chondrite formation region generally have fallen below 600 K, so the dust that is present would contain all of the MOVEs, and therefore, dilute any of the fractionations that had developed.

The necessary conditions to produce these trends can only be achieved by using a narrow range of parameters in the model developed here. The high temperatures in the asteroid belt region can be achieved by a combination of rapid mass transport in the disk leading to the generation of thermal energy in the disk and by maintaining a high dust surface density so that the disk remains optically thick and does not radiate away its energy too quickly. In practice, these two situations are found to go together, as high surface densities provide, locally, a large amount of mass to be transported. This can be seen by looking at the relation between midplane temperature and mass transport ($dM/dt$) given by (Cassen 1994):

$$T_m^4 = \frac{3 \tau G M_\odot}{64 \pi \sigma r^3} \frac{dM}{dt}$$

where $M_\odot$ is the mass of the sun, $G$ is the gravitational constant, $\sigma$ is the Stefan-Boltzmann constant, and $r$ is the location in the disk. The mass transport rate can be written as:

$$\frac{dM}{dt} = 2 \pi r \Sigma V$$

where $V$ is the advective velocity, and is equal to $3\nu/2r$. The optical depth, $\tau$, can be written so that $\tau=\kappa f \Sigma$, where $f$ is the dust-to-gas mass ratio. Remembering that $\nu=\alpha c H=\alpha c^2/\Omega=\alpha(\gamma k T_m/m_{H2})(GM_\odot)^{-1/2}r^{3/2}$, the midplane temperature is found to have the relation

$$T_m^3 \propto \alpha \Sigma^2 f r^{3/2}$$

Thus at a given location of the disk, starting with temperatures above 1350 K at a given distance requires the right combination of $\alpha$ and $\Sigma$. In those cases in which the all material was vaporized out beyond 2 AU, it was found that the disk evolved very quickly, due to the fact that the viscosity was high ($\alpha > \sim 10^{-3}$) or all the mass was concentrated at small heliocentric distances ($\Sigma > \sim 5000$ g/cm$^2$ at 2 AU). As a result, the disk would cool rapidly, meaning that the volatile content of the dust in the disk would change on short timescales, and thus, the planetesimals would have to form rapidly to preserve the fractionated materials from the early, high temperature stages. This generally required coagulation and accretion timescales <$10^5$ years, with shorter timescales limiting radial mixing and therefore leading to steeper depletion patterns. It should be noted that the short coagulation timescales further aided disk cooling as it reduces the dust-to-gas mass ratio ($f$), and therefore lowers the local opacity in the disk, allowing radiation to escape more easily. Higher values of $\alpha$ would lead to faster evolution, and therefore require shorter coagulation and accretion timescales to produce depletion patterns.

Using parameters outside of the necessary range would either not produce planetesimals that exhibit depletion patterns or would limit such planetesimals to very small heliocentric distances. This is demonstrated in Cases 4 and 5 (Figures 6 and 7) where little to no depletion trends develop, as a result of the choices of parameters, and those that do are inside of where chondritic meteorites are expected to form. In Case 4, the same conditions are used as in Case 2, but the coagulation timescale is an order of magnitude longer ($10^6$ years at 1 AU). This longer coagulation timescale prevents the dust from being incorporated into the planetesimals rapidly, and therefore the planetesimals do not preserve as much material from the high temperatures during the earliest stages of evolution. For Case 4, the results are shown after $10^6$ years (1 coagulation timescale), whereas all other cases are shown after 5 coagulation timescales allowing for significant accumulation of material into planetesimals. Allowing Case 4 to continue to run to 5 million would further dilute the depletion trends as the material that would be added to the planetesimals would be unfractionated, due to the lower temperatures that would exist at these later times. Figure 8 shows how this dilution takes place for the planetesimals forming at 2 AU in Case 2. In Case 5, the only significant depletion trend that develops is at 1 AU. As a result of the lower surface density (and thus decrease in mass transport rates and optical depth), temperatures at 2 AU and beyond never exceeded 800 K, meaning all material that was accreted contained the full inventory of elements that condense at higher temperatures. Similar results are found when low levels of turbulence (low values of $\alpha$) are assumed, as the rate of mass transport is too small to lead to significant heat generation. These results are consistent with the findings of Cassen (1996; 2001).

Thus, in order to produce MOVE depletion trends beyond 2 AU in these models, certain conditions must be met. However, these required conditions appear to be inconsistent with other data and constraints on protoplanetary disk evolution and meteorite parent body formation. Around solar mass stars, temperatures above the silicate condensation temperature are difficult to produce and maintain for any significant time at distances beyond 2 AU. There is little, if any, observational evidence to suggest temperatures that high are present in disks around solar mass stars at distances equivalent to the current-day asteroid belt (Woolum and Cassen 1999). This by itself does not pose a problem for this model, as the possibility that the solar nebula evolved differently from those disks that we are able to observe cannot be ruled out. However, the meteoritic data does prove more problematic. The timescales of planetesimal formation that are required here are very short when compared to the half-lives of $^{26}$Al (730,000 yrs) and $^{60}$Fe (1.5 million years). These radionuclides are expected to be major heat sources for the bodies into which they are accreted. Both of these nuclides would condense into the solids at temperatures near ~1350 K and therefore would be incorporated into all solids in this model. It would then be expected that those planetesimals that formed on the timescales given above would differentiate (Ghosh and McSween 1998), rather than preserve pristine nebula material as is done by chondritic meteorites. This effect would vary with planetesimal size, with larger bodies experiencing greater degrees of differentiation as they retain heat more efficiently, and thus could be avoided if chondrite parent bodies were kept <10 km in size for an extended period of time. However, given the large number of planetesimals present, growth of bodies beyond this size is likely to happen relatively quickly (Chambers 2004). In addition, these rapid accretion timescales are difficult to reconcile with chondrules forming 1-4 million years after CAIs (Kita et al. 2000; Amelin et al. 2002).

Not only do these incompatibilities exist, but also, qualitatively, it appears that the planetesimals that would form in these models would be structured very differently than the chondritic meteorite parent bodies. The planetesimals in these models grow through the gradual accretion of more and more volatile materials. Thus, they would start with a core of silicates and metal. As the nebula cooled a bit, the planetesimals would then accrete silicates and metals with MOVEs with condensation temperatures above the ambient temperature. This material would be accreted onto the core of silicates and metal. As the nebula cooled even further to a lower temperature, another layer would be added that contained all the elements that condensed at temperatures above ambient. Within each of these layers that are added to the planetesimals, the elements that would be present would be present at their solar (CI) relative abundances. It is only the *bulk* planetesimal that exhibits a depletion trend. However, the observed depletion trends are measured in centimeter-sized samples of chondrites. It is unclear how the model planetesimals would become homogenized on such a scale. This again, is not by itself absolute proof that this model is inconsistent with meteorite observations, as the details of the accretion process and the post-accretion evolution of the parent body are not considered here. However, no process has been proposed to produce the homogenization required. Also, the fragile organics found within meteorites would have to be late additions to the meteorite parent bodies, and survived the homogenization process (Huss 2004). When these issues are combined with the other incompatibilities raised above, it suggests that chondrite properties were not directly determined by the cooling of the solar nebula from an initially hot state.

The conclusion that the MOVE depletion trend cannot be explained by the dynamical evolution of the solar nebula from an initially hot phase is the opposite of that reached by Cassen (1996; 2001), despite similar models. The reason for this is that Cassen (1996; 2001) did not compare the required conditions to form the MOVE depletions in his models to the constraints on the timing of events in the solar nebula as recorded by chondritic meteorites. This is understandable as it was not until recently that the age differences between CAIs and chondrules and the duration of chondrule formation were measured with uranium isotopic analyses rather than relying on $^{26}$Al (Amelin et al. 2002). The dates that were calculated using $^{26}$Al relied on the uncertain assumption that the isotope was homogeneously distributed throughout the solar nebula. Even with the age confirmations from a second radiometric system, there are uncertainties surrounding these ages as Bizzarro et al. (2004) reported evidence that Allende chondrule formation may have began to occur contemporaneously with CAI formation, though continued for ~1.5 million years.

Cassen (1996; 2001) was aware that the short formation times of the planetesimals required to produce the MOVE depletions in his models imply that they incorporated a significant amount of short-lived radionuclides. Woolum and Cassen (1999) suggested that the differentiation of chondritic parent bodies formed in <$10^5$ years could be avoided if these objects represented those planetesimals that were not accreted into objects >10 km in size. Differentiation is likely restricted to bodies in excess of ~10 km (if formed at such a early stage), and thus if chondrite parent bodies were kept below that size, or grew to that size over an extended period of time, then they may have been able to preserve their nebular signatures.  However, such a scenario is still unable to explain the million year age differences in chondrules and the age differences between CAIs and chondrules.  (It should be noted that Cassen (2001) suggested that the depletion volatile depletion pattern of the Earth could be explained in this model as well, as the planetesimals from which the Earth formed would have evolved in the same manner described here.  This is still possible as these planetesimals need not have avoided differentiation and many would have formed closer to the sun than the asteroid belt distances considered here, where higher temperatures are easier to reach and maintain for significant time in disk models (i.e. Figure 7).)

## Dust Composition

The bulk compositions of chondrites are determined by the sum of their components: matrix, chondrules, and refractory inclusions.  These components have very different origins and compositions, but all formed directly from dust particles in the solar nebula and came together in different proportions to form the chondrites.  Not only do the components have different origins and compositions, but also there are chemical and isotopic variations within the individual components as chondrule compositions vary greatly in a given meteorite and there are a variety of different types of refractory inclusions.

In order to disentangle how these different components achieved their respective compositions, it is necessary to first identify what materials these objects would have been created from.  Here the focus is on matrix and chondrules as that material was likely processed in the chondrite formation region and makes up the bulk (>90%) of a given chondritic meteorite.  Refractory inclusions, such as CAIs, likely formed near the sun, and were then transported outwards to be mixed with chondrules and matrix and accreted into the final meteorite parent bodies (Cuzzi et al. 2003).

During the time of chondrule formation, the dust located in the chondrite formation region would have been a mixture of materials that originated inward, at high temperatures, and diffused outwards ("fractionated") and materials that originated further out, at lower temperatures, and were carried inward by the advective flows associated with disk evolution and gas drag ("unfractionated").  The relative amounts of these two components at a given location varied with time and depended on such factors as the assumed disk viscosity and the disk structure.  Figure 9 shows how this mixture varies in disks of the same structure with various assumed values of $\alpha$.  Plotted are the fractions of dust that was exposed to temperatures in excess of 600 K (approximately the lowest temperature for the MOVE depletion trend) at different times during disk evolution.  This includes dust that began in a location of the disk with T>600 K and that which originated at cooler temperatures, was carried inwards to the warm regions, and then diffused outwards again.

The disk was assumed to have a structure similar to that used in Cases 1-4, where the high starting surface densities in the chondrite formation region allowed for temperatures to exceed 1350 K for $\alpha>10^{-3}$.  No coagulation or accretion of solids into larger bodies was considered; here the focus is strictly on the dynamical evolution of the dust in the disk.  Initially, the dust particles are not mixed and the boundary between the two species (where T=600 K) is seen in Panel A.  As expected, this boundary initially exists at larger heliocentric distances for higher values of $\alpha$, since this leads to higher viscosities and therefore more internal heat generation. As disk evolution proceeds, the disk cools allowing the unfractionated dust to exist at smaller heliocentric distances. Some amount of fractionated material does diffuse outward to mix with the unfractionated materials, however, the amount of fractionated material that remains diminishes with time, particularly for large values of $\alpha$. This is because the fractionated material must survive by diffusing upstream against the net advective flow of the disk. The distance material will diffuse in the disk in a given time, $t$, is given by $x_D=(\alpha cHt)^{1/2}$ (assuming very small particles with $St<<1$). In that same amount of time, the net advective flow of the disk will cause materials to drift inward $x_A=(3\alpha cHt)/2r$. Fractionated particles that originate closer to the sun can survive when $x_D>x_A$, or for a time period of $t<4r^2/(9\alpha cH)$. This means that the amount of time for particles to survive decreases with increasing $\alpha$. Thus those conditions that most easily produce the "hot inner nebula" (high values of $\alpha$) are the same that lead to rapid evolution and advection, causing materials that originated at high temperatures to be accreted onto the sun and leaving materials that originated at low temperatures to be the dominant dust component in the disk. Similar results were found by Cuzzi et al. (2003) who showed that while turbulent diffusion can help prevent the total loss of CAIs over times $>10^5$ years due to gas drag, the surviving particles constituted just small fraction of the solids between 2-4 AU during the expected time of chondrite accretion. Cuzzi et al. (2003) also found that lower values of $\alpha$ provided a better opportunity for these CAIs to be retained in the disk.

Thus, a larger majority of the dust located between 2-4 AU, particularly after a million years of disk evolution (when chondrule formation is thought to begin), likely originated at larger heliocentric distances, and therefore cooler temperatures, and would contain the full complement of moderately volatile elements. This material would have been carried inward from larger heliocentric distances by the net advective flows associated with disk evolution and by gas drag. Even dust that was processed in the inner disk and then diffused outwards may have whatever volatiles it lost recondense on its surface as it entered cooler regions as the volatiles in the gas would diffuse outwards as well.

This means that if chondrules and chondrite matrix formed as products of purely nebular processes instead of planetary, and that the radiometric ages of chondrules do represent a formation period of millions of years, the precursors to chondrules and matrix were likely relatively pristine in terms of their elemental compositions, and therefore CI-like in their elemental abundances. This would be consistent with the suggestion of Alexander (2005) who argued that all chondritic meteorites accreted a matrix that was dominated by a CI-like component. Alexander (2005) argued that the elemental depletions observed in chondrites could be explained by the mixing of such a component with chondrules that lost their volatiles during the heating associated with the chondrule-formation event. Bland et al. (2005) countered that, based on their measurements, matrix in chondrites was not CI-like. However, Alexander (2005) suggested that chondrule fragments may be polluting the matrix and were not recognized. In addition, Alexander (2004) found that the different compositions of chondrules can be explained by considering the kinetic effects of heating and recondensing CI-like materials. This scenario also allows temperatures to be low enough for water ice to accrete with the planetesimals, allowing for aqueous alteration to occur. In this scenario, MOVE depletions would be tied to localized thermal processing of chondritic components.

# A Note on Chemical Kinetics

Ignored in this work has been the potentially important issue of chemical kinetics. Here it had been assumed that the vaporization and condensation reactions were short compared to the dynamical transport times, which likely was not the case for all elements of interest. Fegley (2000) reviewed how some chemical reactions that would produce chondritic minerals were likely kinetically inhibited in the solar nebula, meaning that the amount of time that the reactions needed to go to completion exceeded the expected lifetime of the solar nebula. Kinetics may pose even more of a problem as the transport of a particle within the nebula may decrease the amount of time for a reaction to take place as that particle moves to regions where that reaction was no longer allowed (above the condensation temperature, for example) or was significantly more sluggish than other locations (i.e. in cooler regions). Thus the window in time during which a reaction can take place is likely much less than the lifetime of the solar nebula.

Unfortunately, the reaction rates for all the relevant condensation and evaporation reactions are not known, so it is not possible to include these effects in detail. However, it is expected that even if the reactions were to be slow compared to the timescales of interest, the conclusions of this work would not change. In the case of slow evaporation, as solids entered warmer regions of the nebula, they would retain their volatiles more efficiently than considered here. That would mean that the planetesimals that accreted in the warmer regions of the nebula would have higher volatile contents than considered here, making the depletion trends even shallower, and supporting the conclusion that the MOVE depletion trend could not be directly inherited from a cooling, evolving, nebula.

In the case of slow condensation, it would become possible for solids from the inner disk to diffuse outwards without incorporating volatiles along the way. This would then mean then that volatile-depleted solids would be available to be incorporated into planetesimals in cooler regions of the nebula. A volatile-depleted component (the outwardly diffusing, high temperature materials) would then be mixed together with a volatile-rich component (the pristine materials that never experienced $T > \sim 600$ K). The story would actually be more complex than a two-component model, as the accreted material would actually be composed of a variety of components: a component that never experienced a temperature in excess of 600 K, a component that never experienced a temperature of 800 K, and so on, with the most common component being the more pristine material and the rarest being that which reached temperatures in excess of ~1350 K (because it would have to diffuse outwards the greatest distance and survive in the nebula for the longest period of time). However, as demonstrated in the previous section, the amount of material that originated in the high temperature region of the nebula and was transported out to be accreted in the chondrite formation region is expected to make up only a small fraction of the total mass of solids. Thus, in the absence of localized processes, it would be expected that the material in the chondrite formation region would be dominated by a pristine, CI-like component.

# Discussion and Summary

The objective of the modeling effort presented here was to evaluate whether the depletion of moderately volatile elements observed in chondritic meteorites could be due to the global evolution of the solar nebula and the solids it contained. The thermal evolution of the solar nebula was calculated using the same models used to describe the evolution of protoplanetary disks around young stars, and the dynamical evolution of the solids in the disk were calculated as they grew from dust to planetesimals and were subjected to advective flows, turbulent diffusion, and gas drag migration. A variety of disk structures, levels of turbulence, and coagulation and accretion rates were investigated. It was found that, for a range of parameters, depletion trends of the type observed in chondritic meteorites could be produced in the theoretical planetesimals formed in these models. This is in agreement with the findings of Cassen (1996; 2001).

However, while the depletion trends can be reproduced in these models, there is a finite range of parameters that produce these trends. In particular, rapid accretion of planetesimals are required, which would appear to lead to differentiated bodies rather than chondrite parent bodies, and would be inconsistent with the measurements indicating that chondrules formed over a period of ~2 million years. In addition, it is unclear how these model planetesimals would become as well mixed and homogenized as chondritic meteorites are. As such, it appears that forming the chondritic depletion trends through the gradual accumulation of material in a solar nebula that begins with globally hot inner disk is inconsistent with other observations of meteorites. Instead, the dusty precursors of chondrules and matrix were likely CI-like in composition, which is consistent with the models of Alexander (2004; 2005).

It should also be noted that depletion trends are much easier to develop at smaller heliocentric distances, inside about 1 AU. While Bottke et al. (2006) demonstrated that meteorite parent bodies can be gravitationally scattered outwards from small heliocentric distances to the inner edge of the asteroid belt, it is difficult to imagine that this was the case with all meteorite parent bodies. In particular, the fact that many meteorites were aqueously altered implies that the parent bodies accreted in the presence of water ice, which likely was incorporated into solids outside of ~2.5 AU, as temperatures inward of this location were likely too high for water to exist as a solid. However, this result may still be important for the chemical compositions of the planetesimals that were accreted by the young Earth, as noted by Cassen (2001). This should be looked at more carefully in future work.

As with all theoretical efforts, the conclusions of this work are only as good as the assumptions and treatments that were involved in the model. One of the biggest uncertainties in developing models of disk evolution is identifying the cause for the mass and angular momentum transport. The $\alpha$-viscosity method used here is controversial, as only the MRI has been identified as a process that may generate the kind of viscosity needed to drive disk evolution. However, it is unclear that the MRI will generate this turbulence at the midplane of a protoplanetary disk, and instead, there may be no mass transport in this region. Despite this uncertainty, the particulars of the cause of the mass transport are not critical to this work. What is important is that mass is transported by some mechanism, so that the disk is heated by the loss of gravitational potential energy. In the absence of mass transport, temperatures at the disk midplane would be severely diminished, making it difficult to achieve the high temperatures needed to vaporize all materials in the chondrite formation region. This would only strengthen the conclusion that the evolution of the nebula could not be responsible for the depletion trends observed in chondrites.

This work has also ignored what effects, if any, gravitational interactions with massive clumps or protoplanets may have had on the disk structure and the migration of materials within it. In fact, these types of interactions are possibly responsible for the formation of chondrules as they could generate shock waves in the disk (Boss and Durisen 2005). In addition, the torques may generate a non-uniform surface density distribution and pressure gradients which could lead particles to move *outwards* due to gas drag rather than inwards (Haghighipour and Boss 2003), making it somewhat easier for materials to move from hotter regions of the disk to cooler ones. However, even if this were the case, it is likely that the inward movement of materials from the outer disk would also be increased, providing more CI-like material and averaging out to similar results as those shown here. This should still be looked at closely in future work.

While there are many other ways to improve the model used in this investigation, the simplified scenarios outlined here do provide insight into the general evolution that likely took place in the solar nebula. More complicated scenarios may be envisioned, such as allowing $\alpha$ to vary with time and location in the disk or investigating disk structures that do not have the smooth distributions assumed here. However, it is not obvious why such complications should significantly change the results and conclusions. It is true that we have much more to learn about the evolution of protoplanetary disks, and there certainly are effects that have been neglected that could leave their imprints on the meteoritic record. Future astronomical observations will serve as guides to what other effects must be considered.

## Acknowledgements

This work was done while FJC was funded by the Carnegie Institution of Washington. Discussions with Conel Alexander, Phil Bland, and John Chambers greatly helped the development of this model and manuscript. Discussions with Jeff Cuzzi were instrumental in the early development of this model. A review by Qing-zhu Yin also led to signifancant improvements to this manuscript.

|  | Case 1 | Case 2 | Case 3 | Case 4 | Case 5 |
|---|---|---|---|---|---|
| $\Sigma(1\ AU)\ (g/cm^2)$ | 14000 | 14000 | 14000 | 14000 | 7000 |
| $R_0$ (AU) | 20 | 20 | 20 | 20 | 40 |
| p | -1 | -1 | -1 | -1 | -1 |
| $\alpha$ | $10^{-3}$ | $10^{-3}$ | $10^{-3}$ | $10^{-3}$ | $10^{-3}$ |
| $t_c$ (yrs) | $10^5$ | $10^5$ | $10^5$ | $10^6$ | $10^5$ |
| $t_a$ (yrs) | $10^{-1}$ | $10^3$ | $10^3$ | $10^3$ | $10^3$ |
| Element Diffusion in migrators | N/A | Rapid | Slow | Rapid | Rapid |
| Initial Disk Mass ($M_\odot$) | 0.2 | 0.2 | 0.2 | 0.2 | 0.2 |

Table 1: Parameters used in each of the cases presented here.

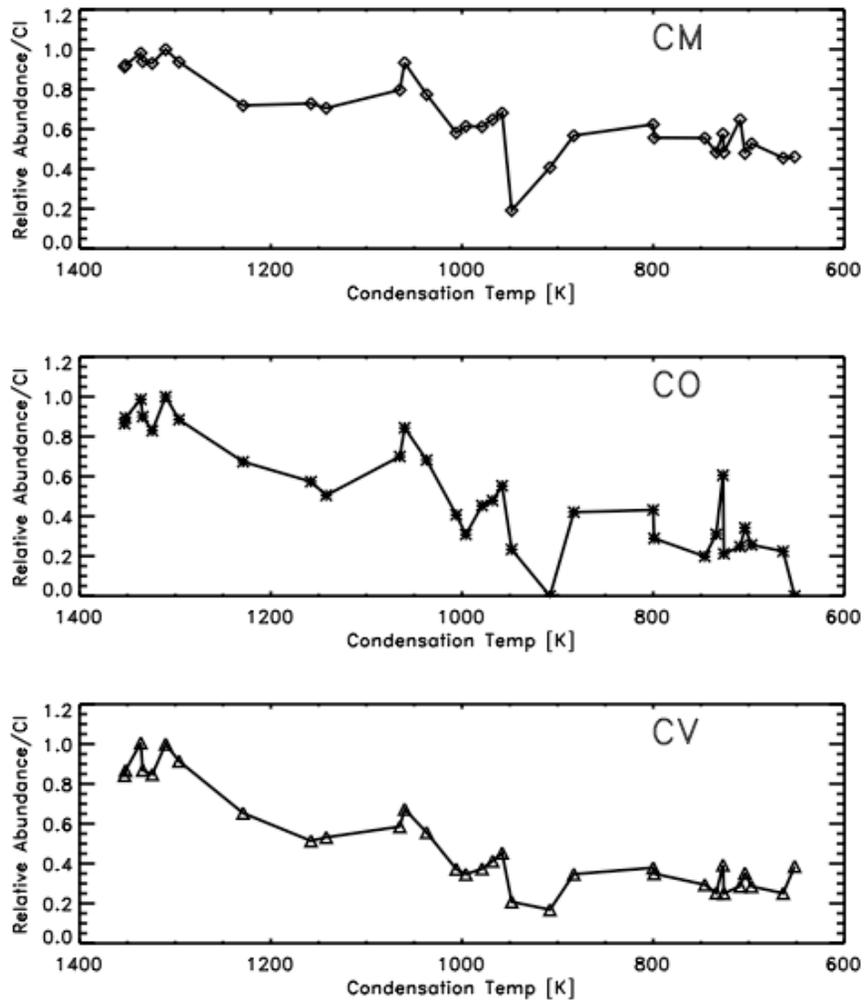

Figure 1: The relative abundances of moderately volatile elements for the CM, CO, and CV chondrites as a function of condensation temperature. The relative abundances are normalized to Si in each type and then divided by the abundances in CI chondrites. Condensation temperatures are taken from Lodders (2003). The plotted elements, in increasing order of volatility, are: Ni, Co, Mg, Fe, Pd, Si, Cr, P, Mn, Li, As, Au, Cu, K, Ag, Sb, Ga, Na, Cl, B, Ge, Rb, Cs, Bi, F, Pb, Zn, Te, Sn, Se, S, and Cd. Data courtesy of Conel Alexander.

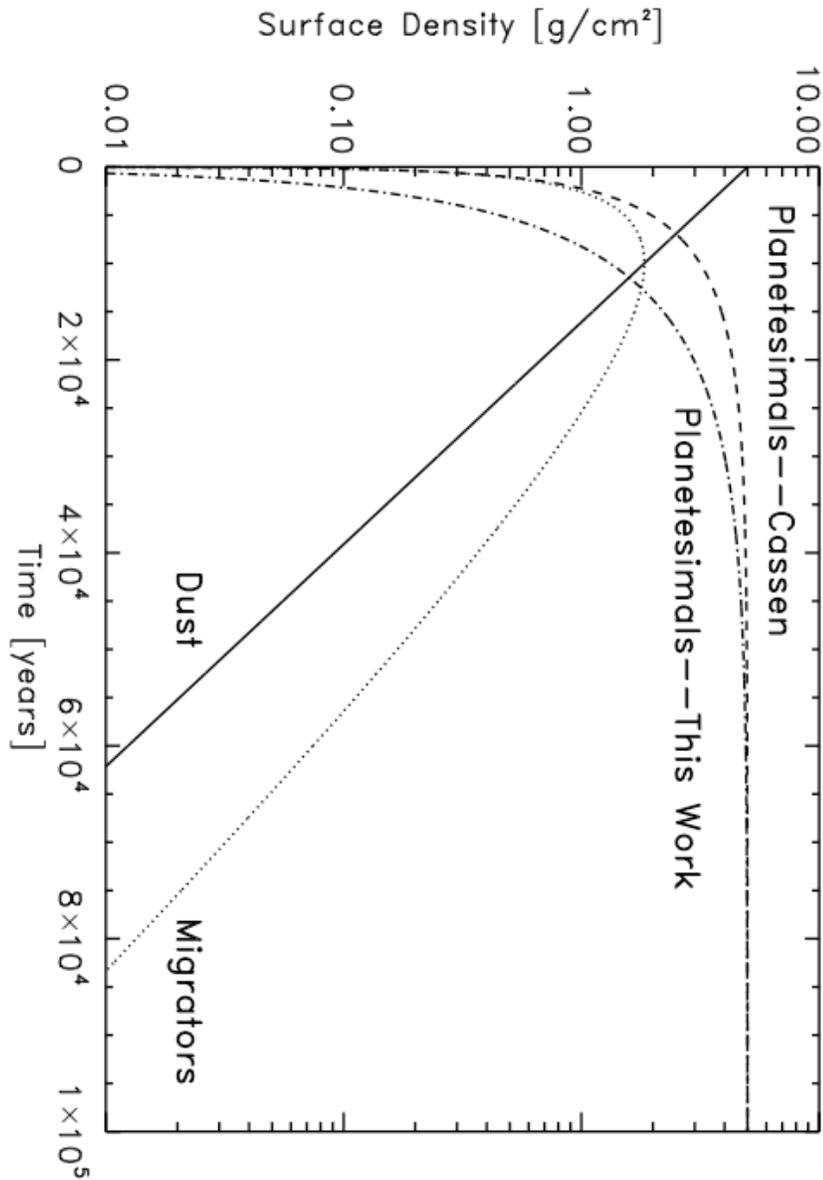

Figure 2: The time evolution of the surface density of dust, migrators, and planetesimals in the single-stage growth model (Cassen) and the two-stage growth model (this work). The dust population (solid line) declines as it is incorporated into larger bodies. In the single-stage model of Cassen, the dust is directly incorporated into planetesimals (dashed line). In the two-stage model used in this work, migrating bodies (dotted line) form which are then incorporated into planetesimals (dash-dotted line). The two-stage model requires a longer period of time for the planetesimals to form.

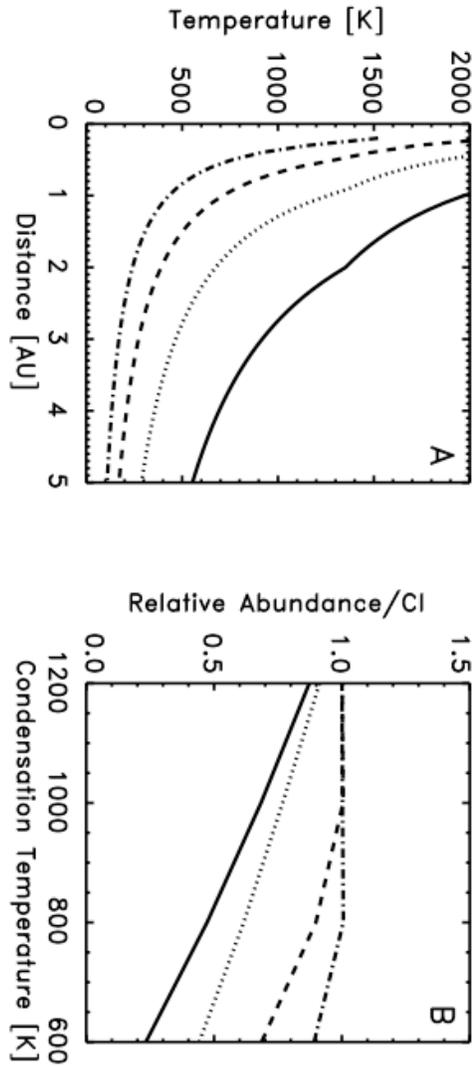

Figure 3: Temperature evolution of the nebula (A) at $10^4$ yrs (solid line), $10^5$ (dotted line) $2.5 \times 10^5$ (dashed) and $5 \times 10^5$ (dash-dotted) and the resulting depletion trends at 1 AU (solid line), 2 AU (dotted), 3 AU dashed) and 4 AU (dash-dotted) (B) for Case 1. This case is similar to those cases considered by Cassen (1996; 2001) in that gas drag migration does not redistribute materials in the nebula. At times beyond $5 \times 10^5$, more material may be added to the planetesimals, but that material would contain the full inventory of MOVEs, and therefore result in the depletion trends growing shallower with time.

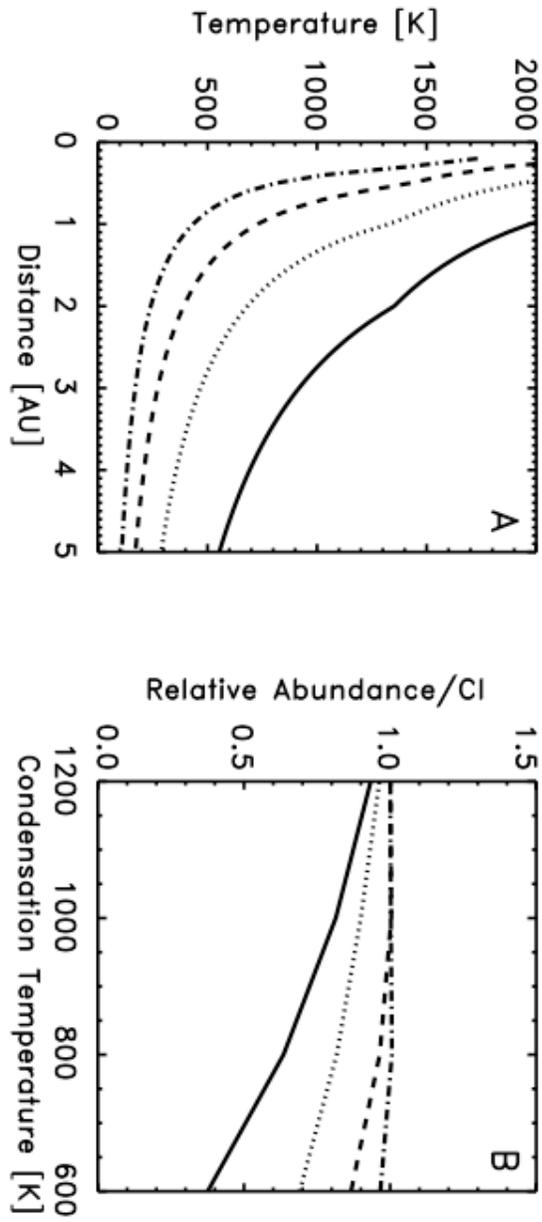

Figure 4: Same as Figure 3, but for Case 2.

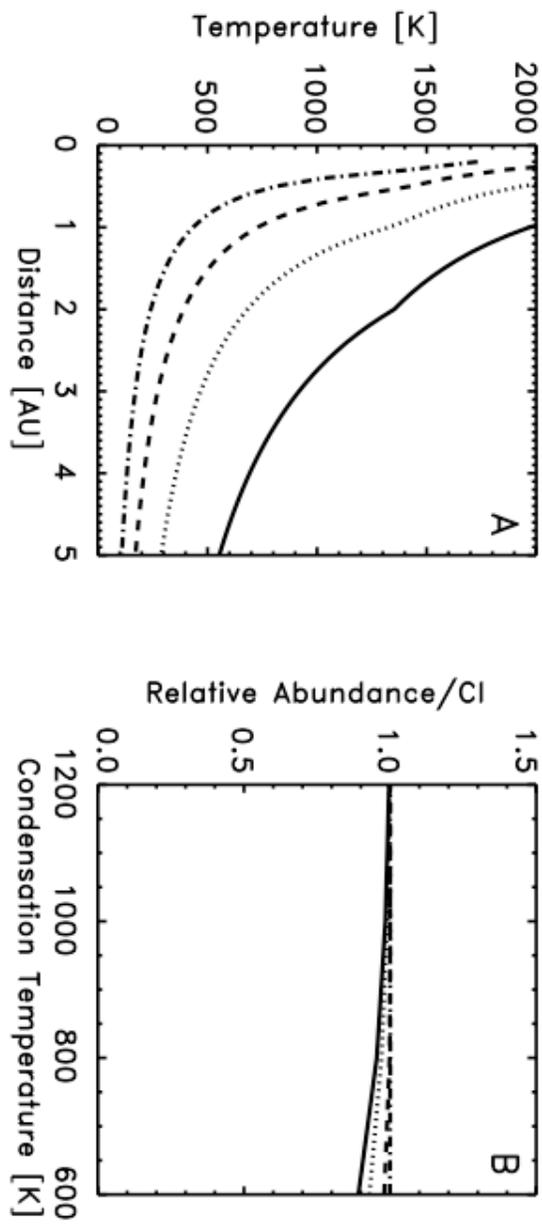

Figure 5: Same as Figure 3, but for Case 3.

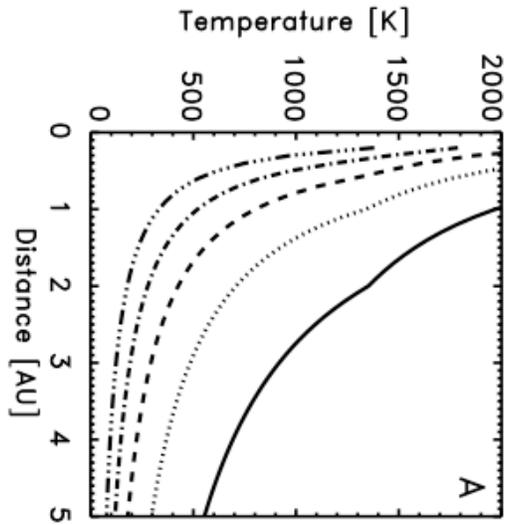

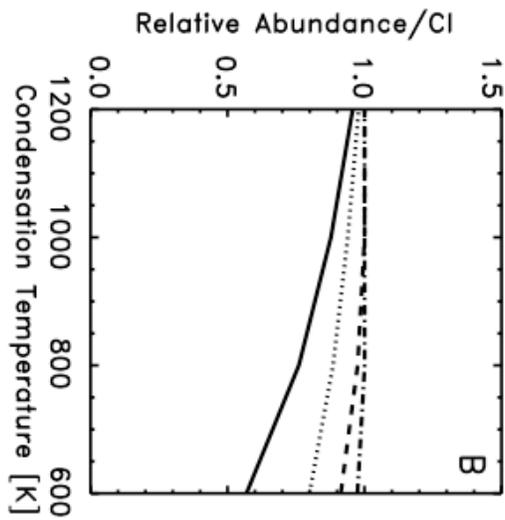

Figure 6: Same as Figure 3, but for Case 4, with the fifth line in Panel A representing the temperature profile after 1 million years. The abundance curves in Panel B are plotted after 1 million years of evolution.

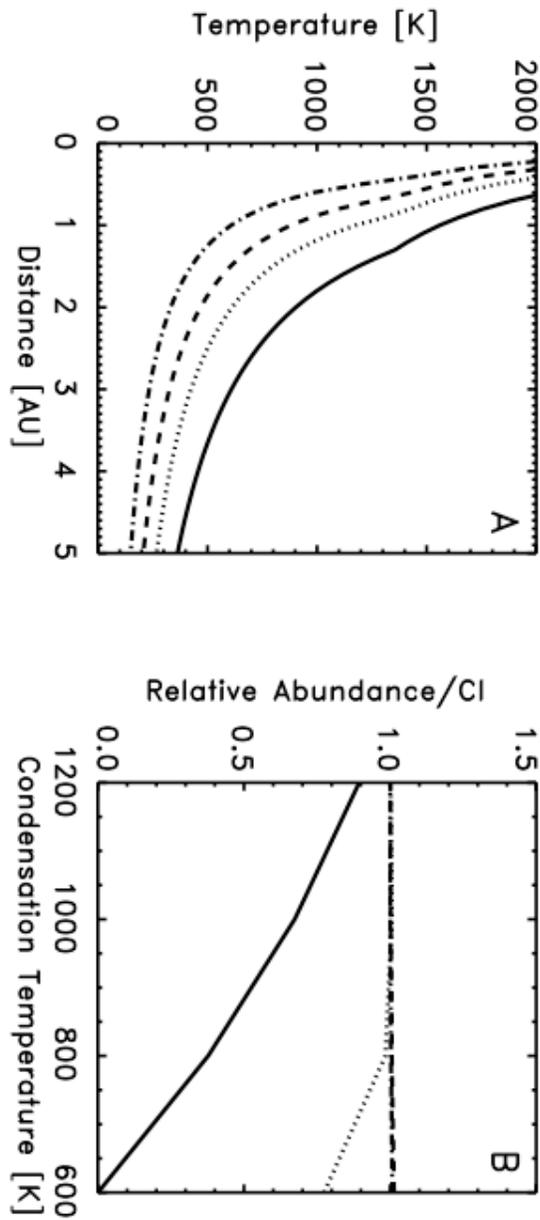

Figure 7: Same as Figure 3, but for Case 5. Significant depletions only develop at 1 AU because temperatures beyond 2 AU never exceeded 800 K in this simulation. As a result, there would be no fractionation of elements with condensation temperatures above 800 K in the planetesimals that formed there.

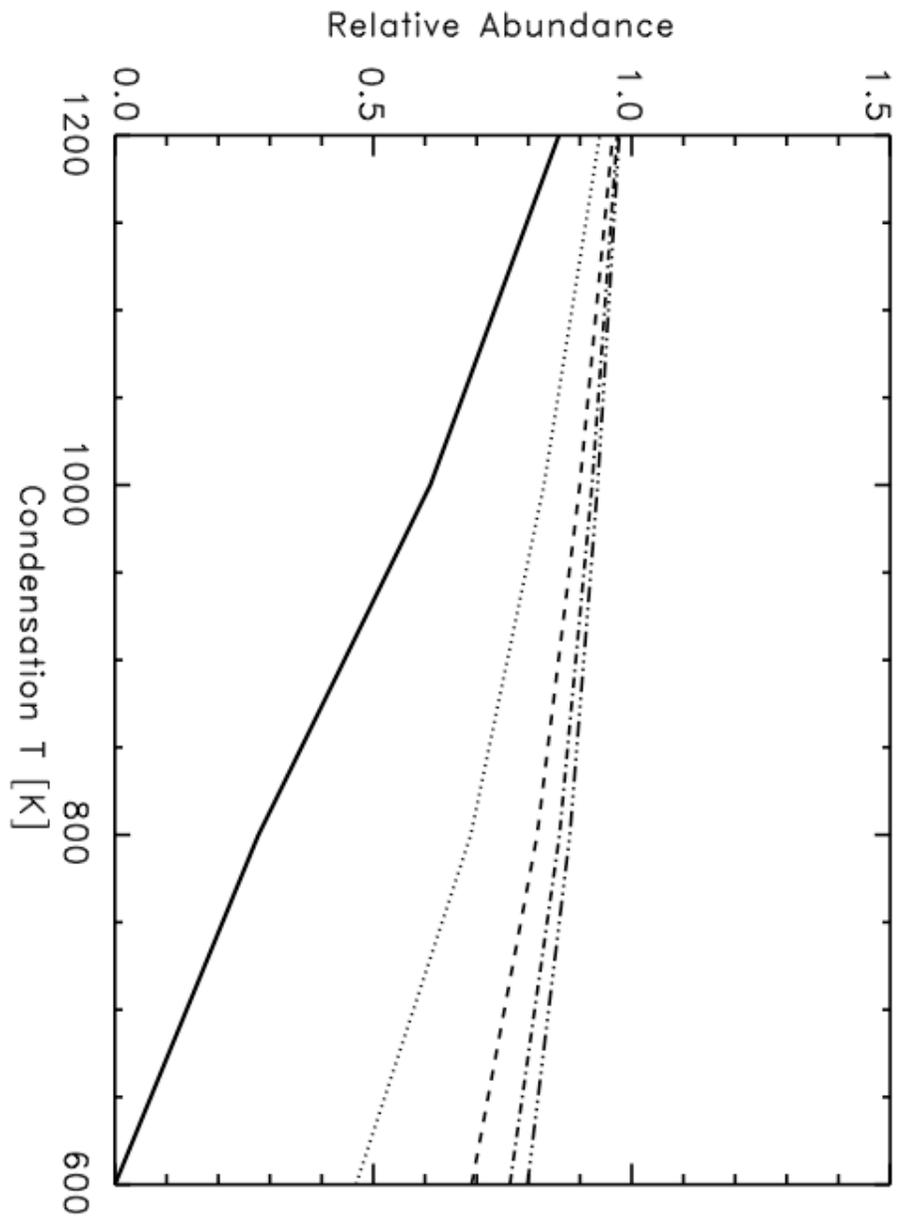

Figure 8: Temporal evolution of the depletion trend for the planetesimals forming at 2 AU in Case 2. The depletion trends are plotted at $10^5$ years (solid line), $2.5 \times 10^5$ years (dotted), $5 \times 10^5$ years (dashed), $7.5 \times 10^5$ years (dash-dotted), and $10^6$ years (dash-dot-dot-dotted). The decrease in the depletion trend with time is due to the continuous cooling of the disk and the incorporation of unfractionated materials.

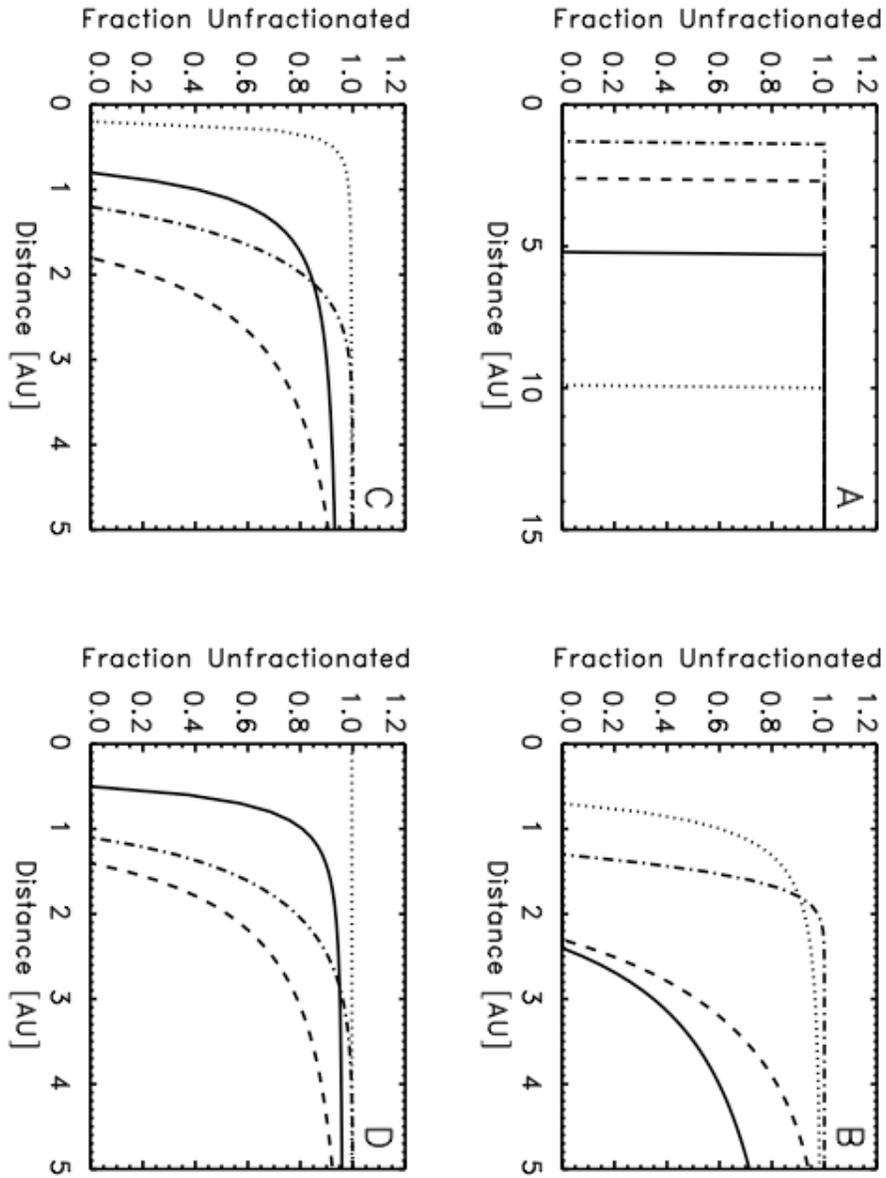

Figure 9: The fraction of "unfractionated" dust (dust never exposed to temperatures >600 K) in the disk at t=0, $10^5$, $5 \times 10^5$, and $10^6$ years (Panels A, B, C, and D respectively). The different curves represent cases for different values of $\alpha$: $10^{-2}$ (dotted), $10^{-3}$ (solid), $10^{-4}$ (dashed), and $10^{-5}$ (dash-dotted). Note the different scale for the X-axis in Panel A.